\documentclass[journal]{vgtc}                
\ifpdf
  \pdfoutput=1\relax                   
  \usepackage{graphicx}                
  \DeclareGraphicsExtensions{.pdf,.png,.jpg,.jpeg} 
\else
  \usepackage{graphicx}                
  \DeclareGraphicsExtensions{.eps}     
\fi%

\graphicspath{{figures/}{pictures/}{images/}{./}} 

\usepackage{microtype}                 
\PassOptionsToPackage{warn}{textcomp}  
\usepackage{textcomp}                  
\usepackage{mathptmx}                  
\usepackage{times}                     
\usepackage{cite}                      
\usepackage{tabu}                      
\usepackage{booktabs}                  

\usepackage{amsfonts}
\usepackage{amsmath}
\usepackage{amssymb}

\usepackage{algorithm}
\usepackage[noend]{algpseudocode}
\usepackage[usenames,dvipsnames]{xcolor}
\usepackage{adjustbox}
\usepackage{tikz}                
\usepackage{enumitem}

\usepackage{hyperref}
\usepackage{comment}
\usepackage{subfigure}
\usepackage{flushend}

\onlineid{0}

\vgtccategory{Research}
\vgtcpapertype{algorithm/technique}

\usepackage[utf8]{inputenc}

\title{Dashboard Design Patterns}

\author{Benjamin Bach, Euan Freeman, Alfie Abdul-Rahman, Cagatay Turkay, Saiful Khan, Yulei Fan, and Min Chen}
\authorfooter{
\item
 Benjamin Bach is with the University of Edinburgh.\\E-mail: bbach@inf.ed.ac.uk.
\item
Euan Freeman is with the University of Glasgow.\\E-mail: euan.freeman@glasgow.ac.uk.
\item
Alfie Abdul-Rahman is with King's College London.\\E-mail: alfie.abdulrahman@kcl.ac.uk.
\item
Cagatay Turkay is with the University of Warwick.\\E-mail: cagatay.turkay@warwick.ac.uk.
\item
Saiful Khan, Yulei Fan, and Min Chen are with the University of Oxford.\\E-mail: saiful.khan@eng.ox.ac.uk, yulei.fan@eng.ox.ac.uk, min.chen@oerc.ox.ac.uk.
}

\keywords{Dashboards, Design Patterns, Data Visualization, Storytelling, Visual Analytics, Qualitative Evaluation, Education}

\CCScatlist{ 
 \CCScat{K.6.1}{Management of Computing and Information Systems}%
{Project and People Management}{Life Cycle};
 \CCScat{K.7.m}{The Computing Profession}{Miscellaneous}{Ethics}
}

\teaser{
    \centering
    \includegraphics[width=1\textwidth]{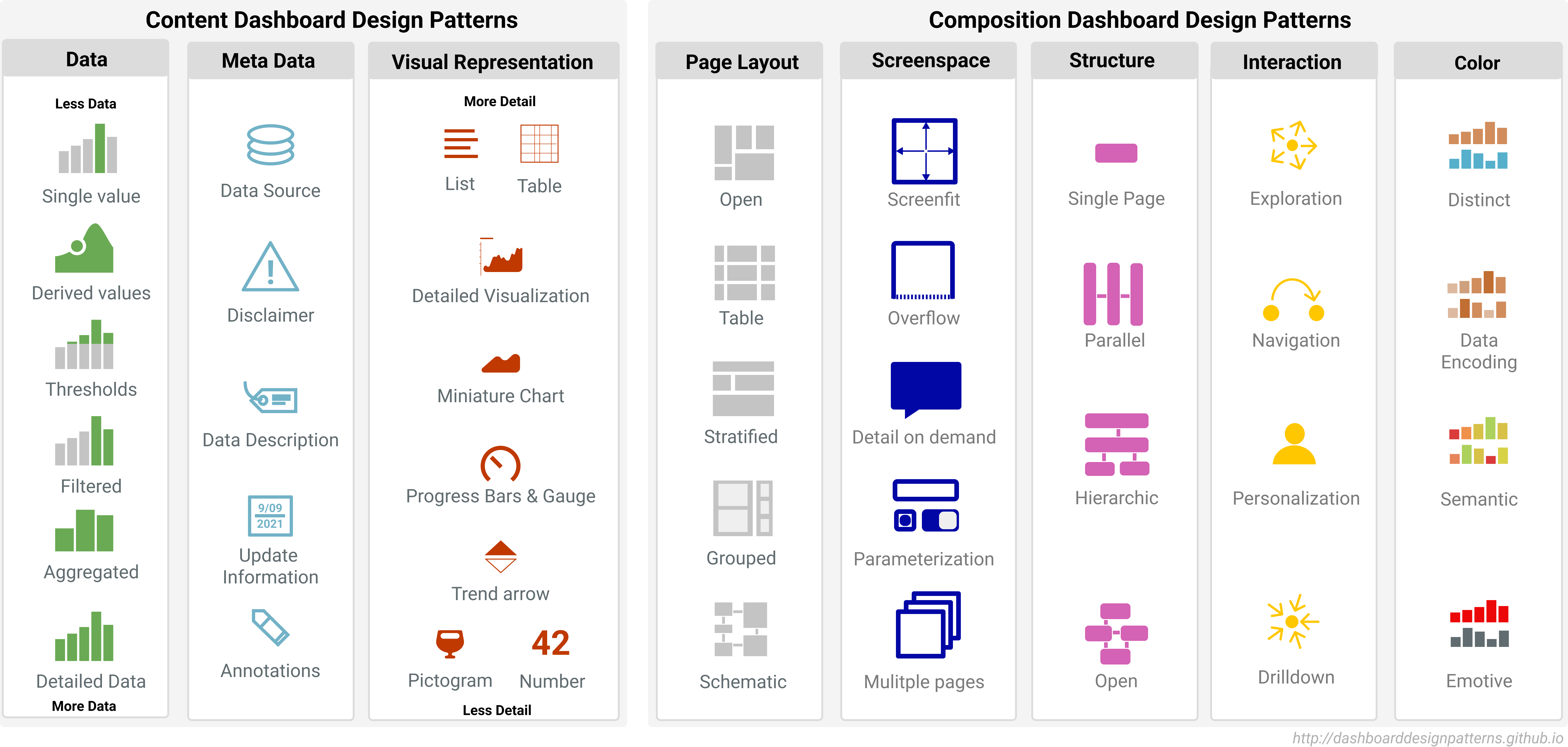}
    \vspace{-3em}
    \caption{Summary of our eight groups of design patterns. Detailed descriptions of each pattern can be found on our website: \url{https://dashboarddesignpatterns.github.io/patterns.html}.}
    \label{fig:patterns}
}

\abstract{This paper introduces design patterns for dashboards to inform dashboard design processes. Despite a growing number of public examples, case studies, and general guidelines there is surprisingly little design guidance for dashboards. Such guidance is necessary to inspire designs and discuss tradeoffs in, e.g., screenspace, interaction, or information shown. Based on a systematic review of 144 dashboards, we report on eight groups of design patterns that provide common solutions in dashboard design. We discuss combinations of these patterns in ``dashboard genres'' such as \textit{narrative}, \textit{analytical}, or \textit{embedded dashboard}. We ran a 2-week dashboard design workshop with 23 participants of varying expertise working on their own data and dashboards. We discuss the application of patterns for the dashboard design processes, as well as general design tradeoffs and common challenges. Our work complements previous surveys and aims to support dashboard designers and researchers in co-creation, structured design decisions, as well as future user evaluations about dashboard design guidelines. Detailed pattern descriptions and workshop material can be found online: \url{https://dashboarddesignpatterns.github.io}%
} 

\vgtcinsertpkg

\usepackage{xspace}
\newcommand{\type}{genre\xspace}
\newcommand{\types}{genres\xspace}

\usepackage{xcolor}

\newcommand{\new}[1]{
{#1}}

\newcommand{\covid}{Covid-19\xspace}

\newcommand{\numdashboards}{144\xspace}
\newcommand{\numpatterns}{42\xspace}
\newcommand{\numtemplatestext}{six\xspace}

\newcommand{\composition}{\textsc{Composition}\xspace}
\newcommand{\content}{\textsc{Content}\xspace}

\newcommand*\inline[1]{~\protect\includegraphics[height=.9em]{#1}}

\newcommand{\linefig}[1]{\protect\raisebox{-.2em}{\inline{figures/icons/#1.png}}}

\usepackage{wrapfig}
\newcommand{\parfig}[1]
{
    \begin{wrapfigure}{l}{.2cm}
    \vspace{-0.4cm}
    \includegraphics[width=.5cm]{#1}
    \vspace{-0.9cm}
    \end{wrapfigure}
    \noindent
}

\begin{document}


\firstsection{Introduction}

\maketitle

Dashboards offer a curated lens through which people can view large and complex data sets \textit{at a glance}~\cite{few2006information,Kitchin2015urbandashboards}. They combine visual representations and other graphical embellishments to provide layers of abstraction and simplification for numerous related data points, so that viewers get an overview of the most important or relevant information, in a time-efficient way. Their ability to provide insight at a glance has led to dashboards being widely used across many application domains, such as business~\cite{few2006information,noonpakdee2018framework}, 
nursing and hospitals~\cite{Mlaver:2017:JQPS,khairat2018impact,wilbanks2014review,bernard2018using,elshehaly2020qualdash,wilbanks2014review}, public health~\cite{Lechner:2014:HCI}, 
learning analytics~\cite{charleer2016creating},
urban analytics~\cite{lee2015cityeye},
personal analytics, energy~\cite{Goodwin2021Energy} and more, summarized elsewhere~\cite{Sarikaya2019Dashboards,yigitbasioglu2012review,few2006information,rasmussen2009business}. These examples, designed mainly for domain experts, have since been complemented by dashboards for public health or political elections, designed for a more general audience and disseminated through news media~\cite{zhang2021mapping} or dedicated dashboard and tracker websites~\cite{dong2020interactive,allison2021interactive, serrano2020political}.

There are many informative \textit{high-level} guidelines on dashboard design, including advice on visual perception, reducing information load, the use of interaction, and visualization literacy~\cite{few2006information,yigitbasioglu2012review,elshehaly2020qualdash,rasmussen2009business}. Despite this, we know little about \textit{effective} and \textit{applicable} dashboard design, and about how to support rapid dashboard design. Dashboard design is admittedly not straightforward: designers have access to numerous data streams, which they can process, abstract, or simplify as they see fit; they have a wide range of visual representations at their disposal; and they can structure and present these visualizations in numerous ways, to take advantage of the large screens on which they are viewed (vs. individual plots that make more economic use of space). These choices can be overwhelming, so there is a timely need for guidance about dashboard design---especially as dashboards are increasingly being designed for a wider non-expert audience by designers without a background in visualization or interface design.

In this work, we analyze the visual design of \numdashboards{} dashboards, to better understand the patterns and design practices used by dashboard designers. By coding these dashboards, we formalized \numpatterns{} design patterns (Fig.~\ref{fig:patterns}) that describe common solutions to design decisions (Sect.~\ref{sec:patterns}). We group the patterns into two high-level groups: \textit{content patterns} that describe what information is shown (\textit{data information}, \textit{meta data}, \textit{visual representation}) and \textit{composition patterns} that describe how components are laid out across one or many dashboard pages (\textit{page layout}, \textit{screenspace}, \textit{structure}, \textit{interaction}, \textit{color}). We then describe six `genres' of dashboard (\autoref{sec:types}) with shared characteristics and common design patterns. These can be related to other genres of information visualization, such as Multiple Coordinated View systems~\cite{roberts2007state} or infographics. Our review and pattern collection help us discuss design tradeoffs (\autoref{sec:tradeoffs}) and discuss possible design frameworks (\autoref{sec:discussion}).

To better understand the role these design patterns and types play in the dashboard design process, we ran a two-week dashboard design workshop (Sect.~\ref{sec:workshop}) with 23 participants. Participants were a mixture of advanced and novice designers and span a variety of backgrounds, both from the academic, public, and private sectors. Participants came with their data and dashboard challenge which we used for group discussions. During the workshop and regular drop-in sessions, most participants worked on dashboard design mockups in the collaborative design platform Figma while others directly designed in tools such as Tableau or Power BI.
The workshop showed that our pattern collection and the associated terminology provided a useful framework to streamline dashboard design for both novice and advanced designers during both, individual and co-design. Discussions revealed challenges in balancing the amount of information shown, designing for a specific audience, and how to best contextualize the information shown. 

Our findings extend prior knowledge about, e.g., dashboard data characteristics~\cite{vazquez2020connecting}, and the intended audience and use of dashboards~\cite{Sarikaya2019Dashboards}. We provide valuable insight into dashboard design, leading to applicable design knowledge that can inform and inspire the creation of future dashboards and dashboard creation tools, help teach dashboard design, and potentially guide structured evaluations into the effectiveness of individual dashboards. A detailed description of all our design patterns, design guidance and the workshop are available online \url{https://dashboarddesignpatterns.github.io}.
\section{Dashboard Design: Background}
\label{sec:relwork}

\subsection{Characterizing Dashboards}

There are many definitions that describe the essential characteristics of dashboards, often exposing opposing views and design guidelines~\cite{few2006information, few2007dashboard, Sarikaya2019Dashboards}. At a high level, dashboards have been described as \textit{tip of the iceberg}, providing a \textit{birds-eye-view} about what a user needs to know~\cite{yigitbasioglu2012review}. Few~\cite{few2006information} highlights four key aspects in describing dashboards as \textit{``a \textbf{visual display} of the \textbf{most important information} needed to \textbf{achieve one or more objectives}; consolidated and \textbf{arranged on a single screen} so that the information can be monitored at a glance''} (emphasis ours). Similarly, Kitchin~\cite{Kitchin2015urbandashboards} echoes that dashboards do not simply reflect data, but are a \textbf{purposefully created} lens through which data must be seen and \textbf{can be engaged with}: \textit{``a dashboard seeks to act as a translator, not simply a mirror, setting the forms and parameters for how data are communicated and thus what the user can see and engage with.''}. Other definitions include the provision of \textbf{metrics} resulting from data analysis, the display of \textbf{dynamic information}, and the ability to provide \textbf{drill-down} capabilities for data exploration~\cite{rasmussen2009business}. 
Elsewhere, Few~\cite{few2007dashboard}, distinguishes between dashboards for monitoring (usually static) and dashboards for analytical tasks (\textit{Faceted Analytical Displays}) which are most \textbf{similar to Multiple Coordinate Views systems}, in that they combine multiple interactive charts and tables. This differentiation is essential because it implies a tradeoff between 
(i) the amount and level of detail of \textit{information} that can comfortably fit a single screen view and (ii) the effort required to access and explore these data through \textit{interaction}.

Dashboards have been reported to serve a range of purposes. They can be designed to support decision making at an executive level (\textit{strategical}), summarize data about departments (\textit{tactical}), and provide information for front-line workers (\textit{operational})~\cite{rahman2017review}. They can also provide consistency with respect to key performance indicators within an organization, help monitor performance, facilitate planning, and support communication~\cite{pauwels2009dashboards}. Through an extensive survey of 83 dashboards, Sarikaya et al.~\cite{Sarikaya2019Dashboards} grouped dashboards into \textit{interactive} dashboards (mostly from BI), \textit{static} dashboards, dashboards for \textit{motivation}, and for \textit{learning} and \textit{personal analytics}. This diverse set of usage goals suggests different solutions that suit the audience, context, and tasks.

Although there is a lack of consensus over what \textit{exactly} should be considered a \textit{dashboard} (versus, e.g., a Faced Analytical Display~\cite{few2007dashboard}), it is clear that dashboards play many roles in their viewers' personal or professional lives. In this work, we are open-minded about what a dashboard can be to give wider insight into the design possibilities.

\subsection{Design Guidelines}
\label{sec:relwork-guidelines}

There is agreement among many case studies and scholars that a dashboard: should \textbf{not overwhelm users}~\cite{yigitbasioglu2012review}; should \textbf{avoid visual clutter}~\cite{few2006information}; should \textbf{avoid poor visual design} and \textbf{carefully chose KPIs}~\cite{rahman2017review}; should align with \textbf{existing workflows}~\cite{faiola2015supporting}; should \textbf{not show too much data}~\cite{janes2013effective}; should have both \textit{functional features} (i.e., what the dashboard can do) and \textit{visual features} (i.e., how information is presented)~\cite{yigitbasioglu2012review}; should provide \textbf{consistency}, interaction \textbf{affordances} and \textbf{manage complexity}~\cite{Sarikaya2019Dashboards}; and should \textbf{organize charts symmetrically}, \textbf{group charts} by attribute, clearly \textbf{separate these groups of charts} and \textbf{order charts according to time}~\cite{bernard2018using}.

Such guidelines can help to inform design at a high level and draw heavily from general knowledge on perception, visualization, and information architecture. For other design decisions, there is less consensus as they require tradeoffs. For example, 
When do we show number values and tables, and when do we show visualizations?~\cite{yigitbasioglu2012review}; 
How much interaction do we include in a dashboard?~\cite{alhamadi2020challenges}; 
How much information to include in a single page?
How to personalize a dashboard?~\cite{vazquez2019tailored}
Does all the content of a dashboard have to fit a single screen size and, if so, how do we deal with different screen sizes? We could not find these questions covered in the literature, and our collection of design patterns aims to complement these high-level guidelines by providing an actionable oversight of solutions used in the wild. Critically for HCI, dashboards have reportedly~\cite{rahman2017review} been rejected by executives on the basis of them not having been involved in the design process, highlighting the need for a user-centered, iterative design process with a shared understanding of concrete design options. 

Closest to our work is a set of visual features identified by Sarikaya et al.~\cite{Sarikaya2019Dashboards}, describing solutions such as multipage dashboards, annotations, and interactive features. While the authors explicitly did not aim for an extensive report on these features, they report on a range of dashboard designs that take inspiration and features from infographics and other storytelling genres. Our dashboard corpus extends theirs with contemporary examples of dashboards and we see our work as an extension of their pioneering work.

\subsection{Design Patterns}
\label{sec:relwork-patterns}

A \textit{design pattern} generally describes a common solution for a recurrent problem. Other than design spaces or taxonomies, design patterns are not exclusive constructs but can be combined and exist independently from each other. Design pattern collections are often used in classrooms and for education~\cite{ideoCards,designwithintend,creativityCards,ObliqueStrategies}.
Pattern collections have been created for visualization, including Card and Mackinlay\,\cite{Card1997}, Chen\,\cite{chen2004toward}, He et al.\,\cite{he2017vizitcards}, Schulz et al.\,\cite{Schulz2013}, and Sedig et al.\,\cite{sedig2016design}. These collections show the breadth of visualization design options and can support designers in making deliberate choices, e.g., about tasks~\cite{Schulz2013}. Design patterns have also been identified for more specialized forms of visualization, including graphical abstracts~\cite{hullman2018picturing}, data comics~\cite{bach2018design}, and sketchnotes~\cite{zheng2021sketchnote}.
Our design patterns for dashboards complement these existing collections and are specific in that they are derived from analysis of \numdashboards~dashboards. Our patterns are purely descriptive in that they capture \textit{existing} solutions and can serve as speculation tools in the design process and discussion, as well as providing a resource for concepts that can inform the structured evaluation of dashboards, a current gap in the literature.
\section{Dashboard Design Patterns}
\label{sec:patterns}

We gathered a corpus of \numdashboards{} dashboards by starting with the 83 dashboards collected by Sarikaya et al.\,\cite{Sarikaya2019Dashboards} but discarding three due to lack of clarity or context. We then added 64 new dashboards found on news websites and personal applications. 36 of these dashboards were related to \covid due to recent public interest. We also included a wide variety of applications (e.g., health and fitness, personal informatics, transport, energy, finance), but given the sheer number and contexts of dashboards, any sampling method is necessary limited.

Three of the authors then created independent coding schemes by qualitatively analyzing the dashboards in the corpus using the constant comparative method~\cite{glaser2017discovery}. Where available, we coded the interactive version of the dashboard (61\%).
These schemes were refined and consolidated with the help of another three additional coders not involved in the initial scheme. Our focus was on coding the structure, visual design, and interactivity of the dashboards---a key distinction between this and prior work, which, e.g., focused on the intentions of the dashboards~\cite{Sarikaya2019Dashboards}. Also, our goal was not to create an exhaustive taxonomy of dashboard design options, but for the first time to describe the user interface building blocks. Our final coding describes \numpatterns{} design patterns, grouped into eight categories, which are divided as follows:

\begin{itemize}[noitemsep,leftmargin=*]
    \item Dashboard \textbf{\content{}} (\autoref{sec:patterns:form}): how the data are abstracted (\textit{Data Information}), what \textit{Meta Information} is included, and what \textit{Visual Representations} are used;
    
    \item Dashboard \textbf{\composition{}} (\autoref{sec:patterns:structure}): the \textit{Page Layout} of components, solutions to fit information into available \textit{Screenspace}, the \textit{Structure} of information across pages, what \textit{Interactions} are supported, and the purposeful use of \textit{Color}.
\end{itemize}

The following section describes these groups of design patterns (\autoref{fig:patterns}). Detailed description for each pattern can be found on our website. Percentages in the following sections indicate the percentage of dashboards where each pattern has been observed; these may not add up to 100\% since the patterns are not mutually exclusive. Numbers preceded by a \#-sign, refer to examples in our collection (see website).

\subsection{Content Design Patterns}
\label{sec:patterns:form}

The \content of a dashboard is made up of individual dashboard elements, the crucial `ingredients' relating to the data and its presentation. We identified three groups of design patterns relating to content: \textbf{data information}, \textbf{meta information}, and \textbf{visual representation} of data. We disregard visual components used purely for decoration or embellishment, e.g., illustrative pictures, dividers, borders.

\subsubsection{Data Information Patterns}

This group of patterns identifies the types of information presented and the extent of abstraction used. We found that information presented roughly ranged from detailed datasets that offer a more complete view of the data, to more abstracted forms that simplify and reduce the amount of information shown (e.g., individual aggregated values, trends). Starting with more complete data, we found \textbf{detailed data sets} \linefig{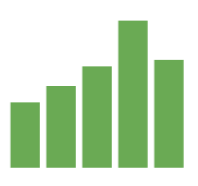} \textbf{(94\%)} which provide the most complete view. \textbf{Aggregation} \linefig{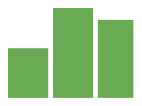} \textbf{(67\%)}, \textbf{filtering}\linefig{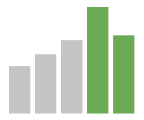} \textbf{(42\%)}, and \textbf{derived values} \linefig{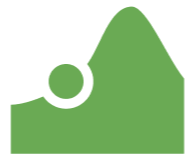} \textbf{(69\%)} provide purposeful summaries of data using summarizing or analysis, e.g., to calculate trends and other measures. \textbf{Thresholds} \linefig{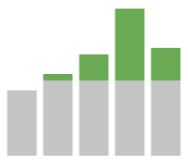} \textbf{(21\%)} \new{indicate states and values that bear some meaning, such as `good' or `bad'. For example, a threshold being passed or data within a specific band or category.} A \textbf{single value} \linefig{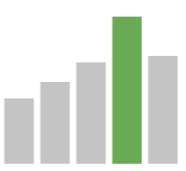} \textbf{(88\%)} of a larger data set an be shown, e.g., the most recent value from a time series.

\subsubsection{Meta Information Patterns}

Meta information patterns capture additional information used to provide context and explanation. In some cases, this is \textit{implicitly} understood from the context the dashboard is used in, e.g., the current date, or data released by a specific organization. \new{In the following, the first percentage values refers to only dashboards with explicit meta information, the 2nd number includes examples where meta data can be implicitly understood from the dashboard context}. We found 9\% of dashboards showed no meta information. 

We found indications of: 
\textbf{data sources} \linefig{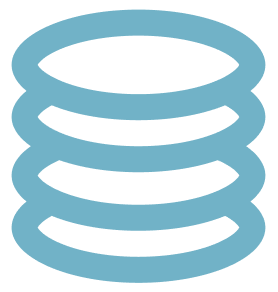} \textbf{(41\%/69\%)};
\textbf{disclaimers} \linefig{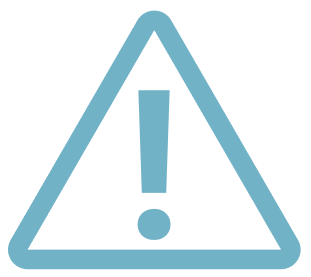} \textbf{(38\%)} informing about data processing and context;
\textbf{data descriptions} \linefig{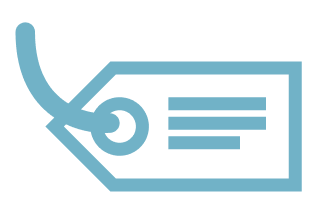} \textbf{(44\%/57\%)} explaining what the dashboard shows;
\textbf{update information} \linefig{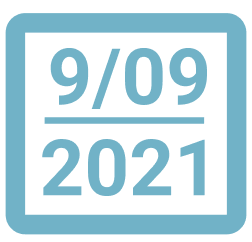} \textbf{(64\%/73\%)} with timestamps; and
\textbf{annotations} \linefig{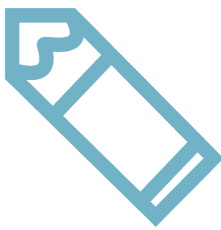} \textbf{(10\%)}, which includes extra graphical embellishments added by the designer to highlight specific points, changes, developments, etc. 

\subsubsection{Visual Representation Patterns}
\label{sec:space:visual}

\begin{figure}[t]
    \subfigure[Pictograms as \textit{data} (\#85, \#37) and \textit{index} (\#102).]{
         \includegraphics[width=1\columnwidth]{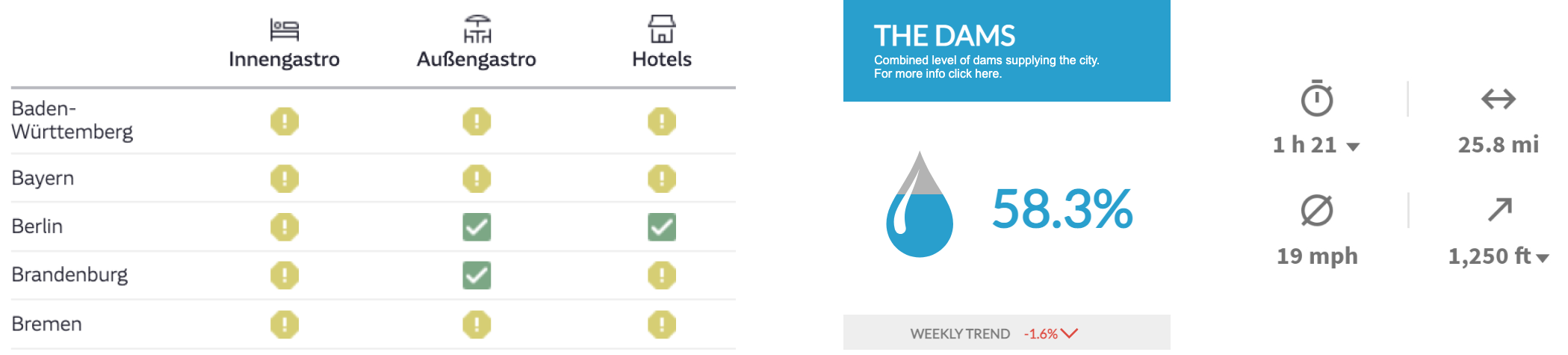}
         \label{fig:data-pictograms}
    }
   
    \subfigure[Gauges and progress bars with accompanying numbers (\#19, \#99, \#16).]{
        \includegraphics[width=.5\columnwidth]{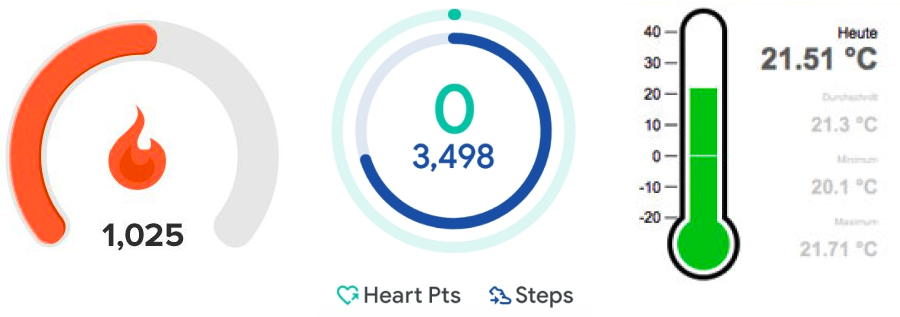}
        \label{fig:data-gauges}
    }
    \subfigure[Miniature charts (\#108, \#5)]{
        \includegraphics[width=.47\columnwidth]{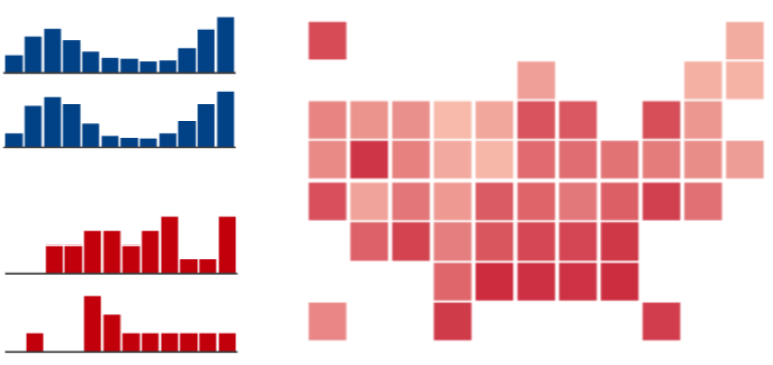}
        \label{fig:data-signaturecharts}
    }
    \subfigure[Table (\#93), data list (\#95), text list (\#93).]{
        \includegraphics[width=1\columnwidth]{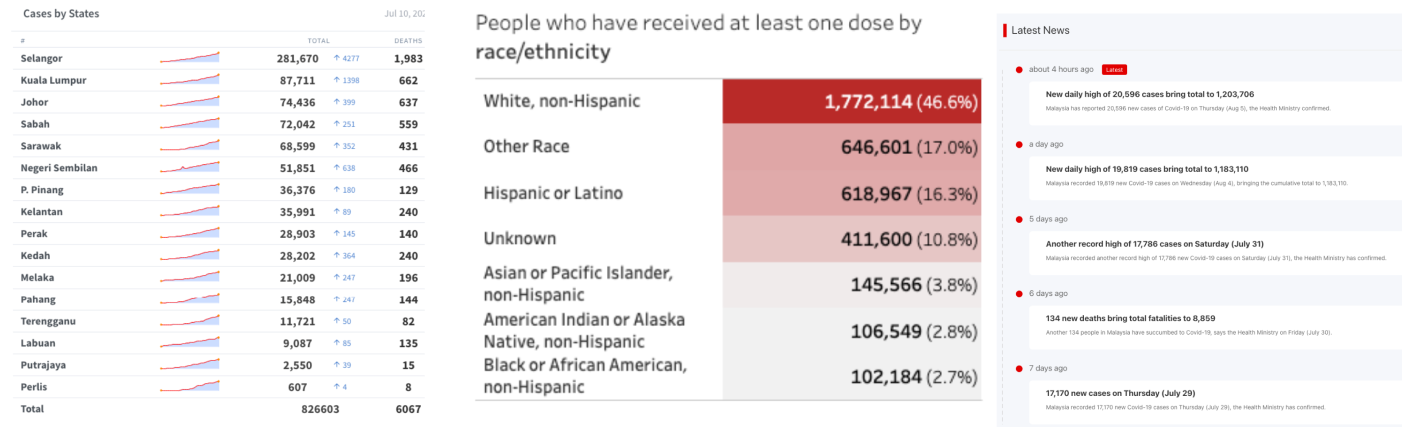}
        \label{fig:data-tables}
    }
    \caption{Examples of visual language used in dashboard to show data.}
    \vspace{-2em}
\end{figure}

We found a wide range of visual representations that, like \textit{Data Information} patterns can correspond to varying degrees of abstraction. 
For example, \textbf{tables \linefig{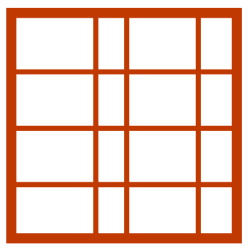} (42\%},  \autoref{fig:data-tables}-middle), 
\textbf{lists\linefig{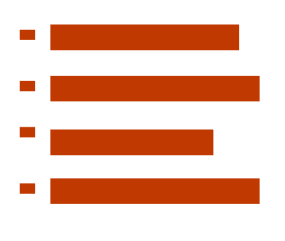} (9\%)}, and 
\textbf{detailed visualizations \linefig{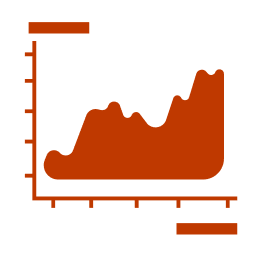} (88\%)} can provide very detailed information and allow viewers to read precise values. Detailed visualizations act as standalone components with proper axis labeling, legends, and resolution. They can span between 1/3 to the full width or height of a dashboard and can be accompanied by numbers and trend arrows (see below). \textbf{Miniature charts \linefig{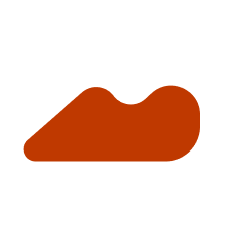} (21\%)}, on the contrary, are small and concise visualizations without axis descriptions, labels, or tickmarks. The idea is to give a quick understanding of a trend, akin to sparklines~\cite{tufte2006beautiful}, rather than allowing the reading of precise values. \autoref{fig:data-signaturecharts} shows examples of miniature charts, all reduced versions of common visual representations. 

More abstract visual encodings include \textbf{Gauges \& progress bars \linefig{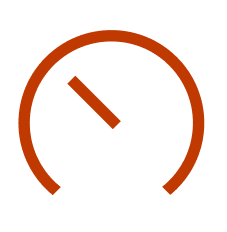} (17\%)} that visualize a single value within its context (e.g., percentage). Specific solutions include semi-circular gauges, linear progress bars, thermometers (\autoref{fig:data-gauges}). Some gauges come with an indication of `critical' ranges, i.e,. a \textit{threshold} \linefig{data-threshold} indicating if values are positive/negative.
\textbf{Pictograms \linefig{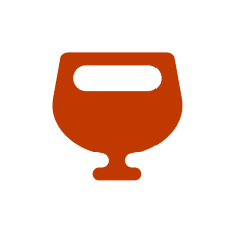} (19\%)} are abstract representations or symbols that illustrate concepts on the dashboard. Their usage can (i)~represent data (\textit{pictogram-as-data}) such as the existence of a value (\autoref{fig:data-pictograms}-left) and a quantity through ``filled pictograms'' (\autoref{fig:data-pictograms}-middle); or (ii)~act as \textit{indices} (\textit{pictograms-as-index}) that designate the type of a data value found close to the pictogram (\autoref{fig:data-pictograms}-right), but not conveying specific data information. 
\textbf{Trend-arrows \linefig{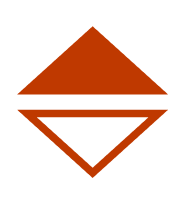} (13\%)} are small arrows pointing up/down and are used to indicate the direction of change in a data value. They can be binary or include variations in slope. Finally, \textbf{numbers \linefig{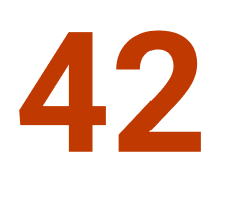} (62\%)} are numerical representations of individual values, placed prominently on a dashboard and mostly used to indicate single values, thresholds, or derived values (\autoref{fig:data-gauges}).

\subsection{Composition Design Patterns}
\label{sec:patterns:structure}

The \composition of a dashboard determines how its individual content components are combined and presented. Dashboards show multiple information elements and their structure and layout on a page are meaningful design decisions. We identified five aspects of composition: 
the \textbf{page layout} of components, 
the methods used to fit the dashboard to the available \textbf{screenspace}, 
the \textbf{structure} of content across multiple pages,
the range of \textbf{interactions} supported by the dashboard,
and the \textbf{color} scheme used by each dashboard.

\subsubsection{Page Layout Patterns}

Page layout patterns describe how \textit{widgets} are laid out and, often, implicitly grouped in a dashboard. There are many layout classifications for information graphics, e.g., for graphical abstracts~\cite{hullman2018picturing} (\textit{linear, circular, zig-zag, forking, nesting, parallel, orthogonal, centric, free}), infographics~\cite{bach2018design} (\textit{large panel, annotated, tiled, grouped, grid, parallel, network, branched, linear}), or sketchnotes~\cite{zheng2021sketchnote} (\textit{freeform, grid, radial, linear}). Here we identify the most prevalent page high-level layout patterns found in our dashboard corpus.

\begin{figure}[h]
\vspace{-.5em}
\includegraphics[width=1\columnwidth,clip,trim=0 5mm 0 0]{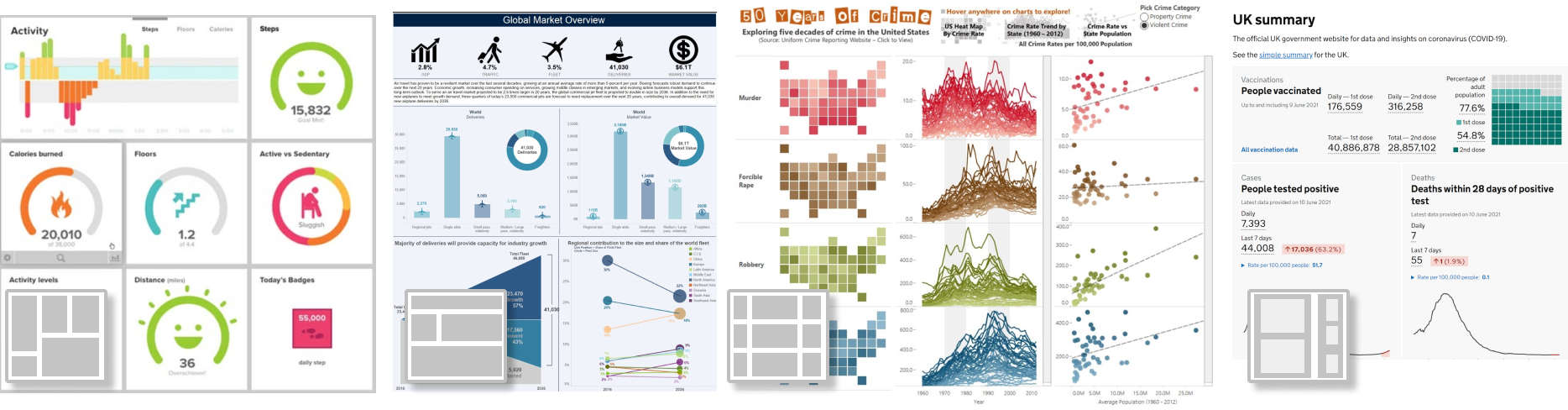} 
\vspace{-1.5em}
\caption{Examples of different dashboard page layouts: \textit{open} \#19, \textit{stratified} \#11, \textit{table} \#5, \textit{grouped} \#88.}
\label{fig:layouts}
\vspace{-1em}
\end{figure}

When describing page layout patterns, we define a \textit{widget} as individual layout components, usually figuring one or a combination of visual representations, a title, and possible meta information. 
Many dashboards group information and their visual representations (\autoref{sec:space:visual}) hierarchically and it is notoriously hard to create a clear definition of grouping in any visual composition. For example, the first dashboard in \autoref{fig:layouts} combines \textit{numbers}, \textit{gauges}, and \textit{pictograms} together to form a specific form of complementary information grouping; the pictogram, number and gauge all relate to the same piece of information or at least can be understood as a semantic-visual unit aiming for a specific task. With our page layout patterns, we consider how these higher-level widgets are organized in a single page, identifying the prominent layout decision used to group a potentially large set of content components. 

\textbf{Open layouts \linefig{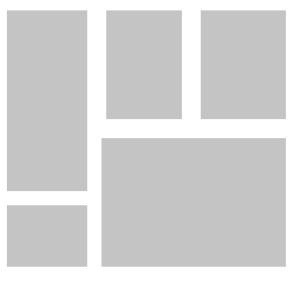} (22\%)} place widgets (possibly of different sizes and aspect ratios) in an open way without apparent specific rules. Often widgets are aligned on a grid (\autoref{fig:layouts}-\#1) following classical design guidelines. There is no strong semantic associated with the location and adjacency of widgets and each widget seems to have equal importance.
\textbf{Stratified layouts \linefig{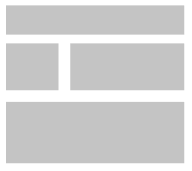} (49\%)} present widgets in a top-down ordering. A stratified layout can be used to emphasize information on the top (\autoref{fig:layouts}-\#2) over other information.
\textbf{Table layouts \linefig{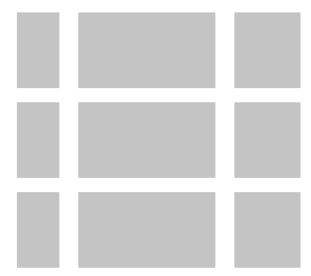} (19\%)} align widgets into semantically meaningful columns and rows. They can be used to repeat information and visual encoding, e.g., across different facets or data items (\autoref{fig:layouts}-\#3). Table layouts make it easy to retrieve and relate information.
\textbf{Grouped layouts \linefig{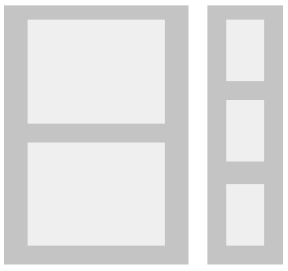} (33\%)} group two or more widgets with a specific relation, in many cases labeled by a common title. Grouping can be achieved through e.g., Gestalt laws of proximity or closure. Finally, \textbf{schematic layouts \linefig{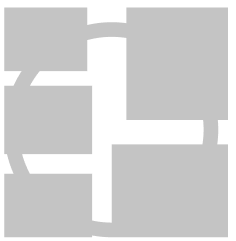} (1\%)} place widgets in some schematic relationship such as a physical-spatial layouts (\#59), networks (\#42), or possibly process-workflows.

We emphasize that none of these page layout patterns are exclusive and combinations are common. For example, the second dashboard in \autoref{fig:layouts} shows a \textit{stratified} layout (pictograms on the top, visualizations on the bottom), combined with an \textit{open} layout. Similarly, the fourth dashboard in the same figure combines a \textit{stratified layout} with a \textit{grouped layout}, emphasizing the key indicators on top.


\subsubsection{Screenspace Patterns} 

Screenspace patterns describe solutions used to fit content onto a single screen. We call these screens \textit{pages} as content can be split across multiple pages or overflow a screen. However, at any given time, only a single page is visible to the viewer. 

\textbf{Screenfit \linefig{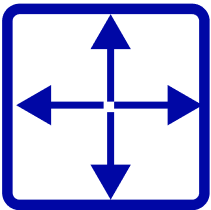} (44\%)} means that all content of a page fits the screen without the need for interactions like scrolling or tooltips. This is the standard solution for concise static dashboards. 
\textbf{Overflow \linefig{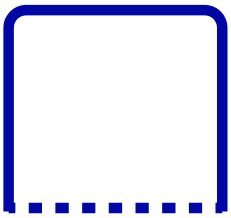} (22\%)} allows a page to be larger than the screenspace available. Overflows are usually explored through vertical scrolling and there is potentially no limit to the size of the overflow. 
\textbf{Detail on demand \linefig{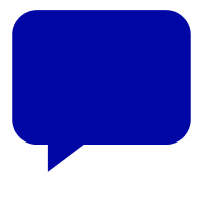} (47\%)} shows extra content on, e.g., mouseover through tooltips, open pop-ups on clicking buttons or widgets, or giving access to more data without needing to fit alongside other content. 
\textbf{Parameterization \linefig{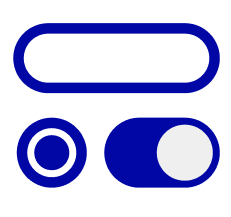} (52\%)} is a further means of \new{defining the information visible, e.g., showing more data, or specify filters on a dataset.} Parameters can be set through sliders, checkboxes, or drop-down menus. Other than overflow\linefig{pagination-scroll} and detail-on-demand\linefig{pagination-tooltip}, parameterziation can show potentially very large and multifaceted data sets, but require manual specification and can typically only show one state at a time. \new{Both, detail-on-demand and parameterization help making decisions about the use of screenspace---notably by hiding data and revealing it back through interaction such as exploration or drilldown, discussed in \autoref{sec:patterns:interaction}).}
\textbf{Multiple pages \linefig{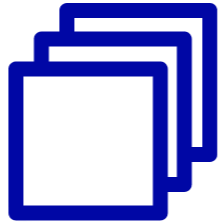} (42\%)} splits content across multiple separate pages which are obtainable through navigation patterns (see \autoref{sec:patterns:interaction}).

\begin{figure}[h]
    \vspace{-.5em}
    \includegraphics[width=1\linewidth]{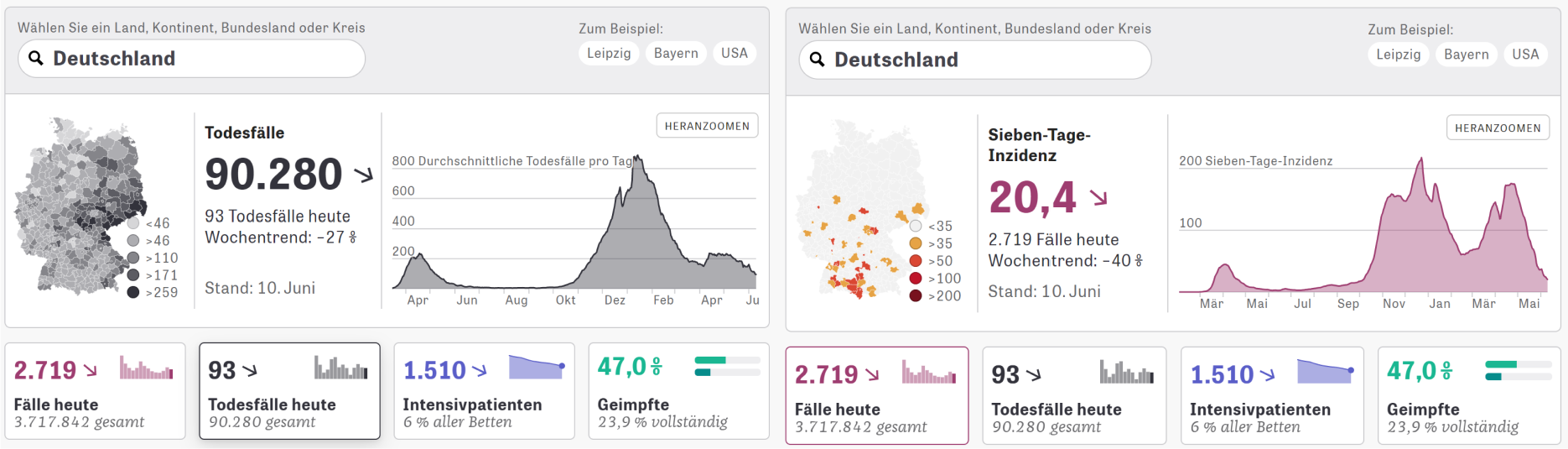}
    \vspace{-1.5em}
    \caption{Example of a tabbed dashboard (\#83). Tabs show a similar view with different data (e.g., \covid deaths and cases here); multiple pages allows a screenfit design through reduced information on any one page.}
    \label{fig:pagination-tabs}
\vspace{-2em}
\end{figure}

\subsubsection{Structure Patterns} 

Structural patterns capture relations between multiple pages \linefig{pagination-multiple} of a dashboard, if there are multiple pages. In the most simple case, a dashboard has a \textbf{single page} \linefig{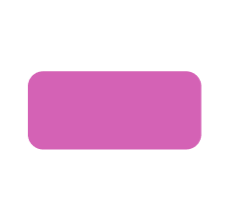} \textbf{(61\%)}, which can imply the need for \textbf{screenspace} patterns such as overflow, details-on-demand, or parameterization to fit the desired content into a single page or screen. Multiple pages can exist through relationships between these pages. 
A \textbf{parallel structure} \linefig{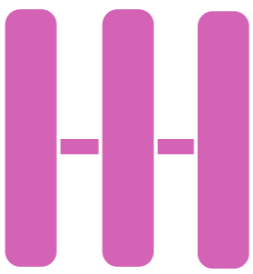} \textbf{(16\%)} can imply repetition of the layout, data, and visual representations, e.g., across different courses in a university, or departments in a company. A parallel structure can be combined with hierarchical, e.g., in case of geographic regions which have a political hierarchy but otherwise show similar information. 
\textbf{Hierarchical structures} \linefig{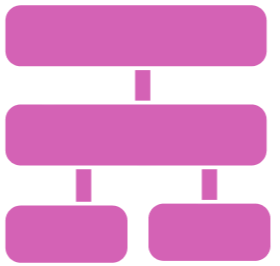} \textbf{(19\%)} are used for drill-down and can result in a series of pages, each gradually showing more detail. 

\textbf{Open structures} \linefig{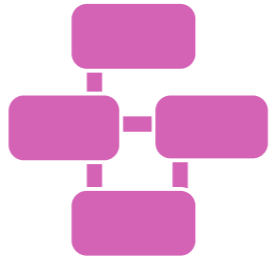} \textbf{(8\%)} capture other kinds of structural relationships.

\subsubsection{Interaction Patterns}
\label{sec:patterns:interaction}

This pattern group describes common interaction approaches found within dashboards. Interaction routines manifest themselves through interactive data entities, i.e., \textit{data as the interface}, user interface elements, as well as window-level interactions. The patterns we highlight in this group refer to common \textit{roles} that interaction could play in dashboard use, expressed through specific user interface components. These are defined broadly, to identify general usage patterns and how they can be implemented through dashboard designs. Only 4\% of dashboards in our corpus had no interaction. We found four major uses of interaction in dashboards, supported through an often overlapping range of UI components, such as tabs, sliders, drop-down menus, etc. 

\textbf{Exploration \linefig{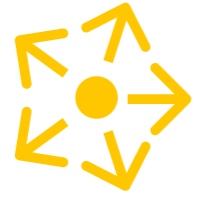} (89\%)} interactions allow users to explore data elements, \new{obtain new data, look at data differently} and explore relations between data. Exploration can take on many forms, including \textit{brushing + linking (18\%)}~\cite{buja1991interactive} that link data representations across different views. Details of individual or groups of items can also be revealed through interactive features, e.g., pop-ups with details, a table enlisting data features, etc (71\%).
In contrast, \textbf{Drilldown \linefig{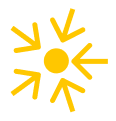} (55\%)} allows viewers to find or focus on specific data, \new{i.e., features a user is interested in during a specific task. Examples include} searching for particular data values or applying filtering criteria so that only relevant information remains. These interactions are typically facilitated by user interface elements, like text fields, drop-down menus, radio buttons and checkboxes.
\textbf{Navigation \linefig{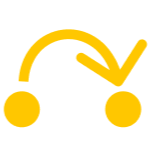} (76\%)} enables navigation between pages of a dashboard or screen \new{features of the dashboard created by the dashboard designer.} Navigation can be supported through scrollbars, navigation buttons, page tabs,  hyperlinks, etc. Eventually, interactions for \textbf{Personalization \linefig{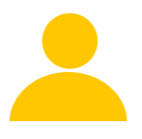} (23\%)} allow viewers to redefine and reconfigure the information shown within a dashboard \new{based on personal preferences and task needs}. Interactions can add new visual representations, e.g., choosing a new data feature to be visualized, resize, or reorder the existing encodings within the dashboards, leading to more bespoke dashboard configurations. 

\subsubsection{Color Patterns}

Color is an important visual variable in visualization. While it can be used for different purposes in dashboards and comes with cultural implications, we examined the use of color at dashboard-level---i.e., across multiple widgets and views.

\textbf{Shared color schemes \linefig{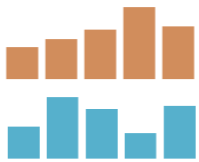} (35\%)} give unique recognizable colors to groups or facets in the data. This helps maintaining consistency and familiarity across throughout the entire dashboard (e.g., \autoref{fig:pagination-tabs}). 

\textbf{Data Encoding \linefig{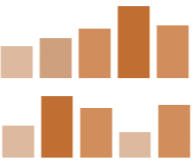} (80\%)} color schemes use colors primarily as a visual variable to encode categories or scales within the data, e.g., displaying values with color on a choropleth map (\#5 in \autoref{fig:layouts}).
\textbf{Semantic \linefig{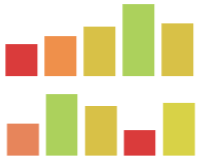} (26\%)} colors indicate specific good-bad semantics, such as the traffic light schemes to indicate status or statuses of patients in a health related dashboards. Semantic coloring is often in conjunction with gauges and progressbars for indicating multiple meaningful thresholds.
\textbf{Emotive \linefig{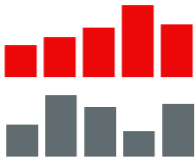} (6\%)} color schemes can add aesthetic strength and develop an emotive response in viewers~\cite{kennedy2018feeling}. This use of color within dashboards seems particularly common in dashboards that resemble infographics (see next section, e.g., \#19, \#60).

\section{Dashboard Genres}
\label{sec:types}

During our analysis of the dashboard corpus, we realized \numtemplatestext dashboard types that can be seen as design \textit{\types}, because they share characteristics, combinations of design patterns, contexts, or specific goals. After describing an initial set of \types{} and creating a respective codebook, we again applied structured coding to cover all the dashboards in our collection. Coding was done by two independent coders coding each dashboard independently. We disagreed on 32 dashboards (22\%) but resolved all disagreement in discussion, refining the respective genre definitions. Similar to the storytelling genres by Segel and Heer~\cite{segel2010narrative}, we focus on the \textit{existence} of a genre and the design tradeoffs, rather an exclusive taxonomy. These \types can invite discussion about how a specific pattern has been put into practice through dashboard design, can be used in design exploration, and can inform discussion about the `right' dashboard design for a given context. 

\parfig{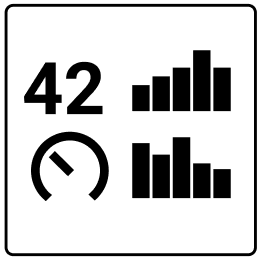}\textbf{Static Dashboard (21\%)---}By static dashboard, we refer to the traditional notion of a dashboard as a static (\textit{no-interaction}), single page\linefig{struct-single}, screenfit\linefig{pagination-screenfit} display of information. \new{Static dashboards often feature concise information and representation such as single-values\linefig{data-simple}
and derived values\linefig{data-derived}, 
miniature charts\linefig{data-signature}, 
arrows\linefig{data-arrows}, and
numbers\linefig{data-number}} (\autoref{fig:layouts}--left (\#19)). We did not find many examples of classic static dashboards, which we attribute to the fact that contemporary dashboards are digital and it is easy to support interaction and drill-down tasks through more complex structures. Likewise, there is a large range of display sizes from desktop computers and tablets, to mobiles that encourage responsiveness (e.g., use of \textit{overflow} to utilize screenspace or a multi-page structure).

\parfig{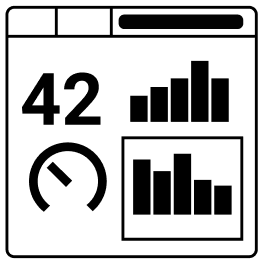}\textbf{Analytic dashboards (73\%)---}This \type is what Few would call a \textit{Faceted Analytic Display}~\cite{few2007dashboard}. We see parallels to the concept of Coordinated Multiple Views (CMV)~\cite{roberts2007state}, while there are clear differences between both concepts: by definition, the focus in CMVs is on coordination and individual views react to interaction in another view to give complementary views for the task. In contrast, a dashboard can have multiple views \textit{without} any shared data, information, or interaction; each view can be entirely independent and not required for a common task. Analytic dashboards generally use complete \textit{visualizations}\linefig{data-visualization} and tables\linefig{data-table} to show larger and more detailed data sets\linefig{data-complex}. Many of these views are fully interactive, providing for exploration\linefig{int-exploration}, navigation\linefig{int-navigation}, and drilldown\linefig{int-filter_focus}. These dashboards can also provide \textit{parameterization}\linefig{pagination-parameterization}, and \textit{tabs}\linefig{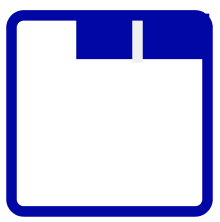} or other \textit{linking} mechanisms to switch between \textit{multiple pages} of the dashboard. Importantly, these dashboards generally do not use \textit{overflow} \linefig{pagination-scroll} pagination, since scrolling complicates comparing visualizations.

\parfig{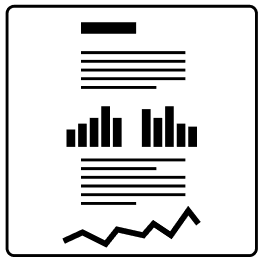}\textbf{Magazine Dashboards (2\%)}---Many dashboards relating to \covid, climate change, politics, etc were created by news agencies and similar media outlets. These dashboards are found as integral part of journalistic articles and resemble visualizations of the \textit{magazine} genre~\cite{segel2010narrative}. The text provided alongside visualizations goes beyond basic meta information to provide additional commentary and storytelling about the data. These dashboards are often broken into several pages and have an \textit{overflow}\linefig{pagination-scroll} use of screenspace on a \textit{single page}\linefig{struct-single}, with visualizations positioned at appropriate points in the text to tell a story about what the data shows. \new{Magazine dashboards can feature very detailed visualizations\linefig{data-visualization} and tables\linefig{data-table}.} The Economist \covid tracker (\#131) is an example that provides viewers with a snapshot of \covid cases and deaths across Europe, with tables, timeseries, trend lines and spike maps interleaved with narrative text. In addition to regular visualization updates, written content is also frequently updated as the `story' changes, e.g., responding to emerging trends, the effects of vaccination, etc. These dashboards naturally require more effort to design and maintain; whilst visualizations may update automatically as the data changes, editorial oversight is necessary to ensure the story remains consistent with the changing data and its visual representation.

\begin{figure}
    \centering
    \includegraphics[width=.95\columnwidth]{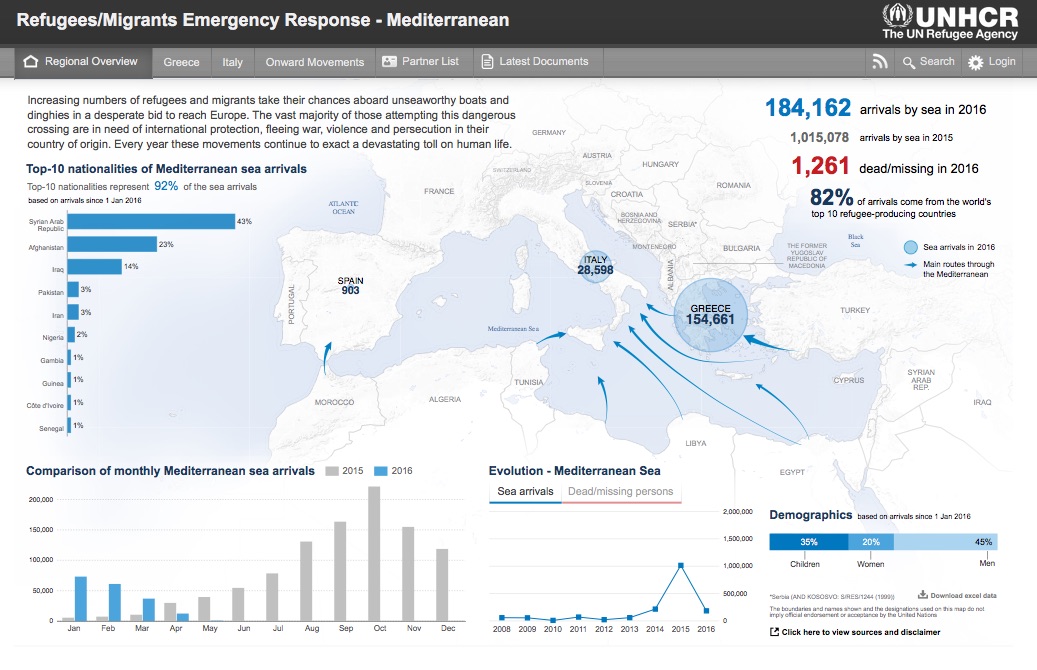}
    \vspace{-1.5em}
    \caption{Example of an infographic style dashboard (\#74).}
    \label{fig:infographic}
    \vspace{-2em}
\end{figure}

\parfig{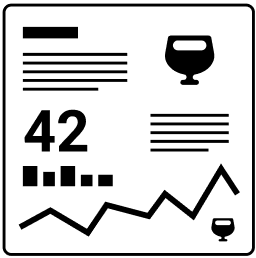}\textbf{Infographic Dashboards (6\%)}---We also found examples of dashboards similar to infographics that include decorative graphical elements and other non-data ink shown alongside data representations. Similar to magazine dashboards, they use non-data media to annotate and embellish data. For example, \autoref{fig:infographic} shows an infographic style dashboard that uses text, annotations and other embellishments to enhance data presentation and, in turn, help the data to convey a story. Other examples include less artistic content, while being presented more as posters (e.g., \#33, \#34). \new{Infographic dashboard can feature catchy gauges\linefig{data-gauge}, pictograms\linefig{data-pictogram}, numbers\linefig{data-number}, and detailed visualizations\linefig{data-visualization}.}

Infographic dashboards were mostly used to represent static datasets; e.g., presenting snapshots of key data on a monthly or yearly basis. Often these infographics exceeded the vertical screen-space and could be explored through scrolling \textit{(overflow})\linefig{pagination-scroll}. The artistic content of infographic dashboards may require additional design time and chosen annotations and embellishments will be tailored to particular data points, so are less suited for dynamic dashboard use where data changes often. These dashboards may thus have a different intended use, with an audience expected to discover them over a longer period of time, rather than checking for frequent updates.
\newline

\parfig{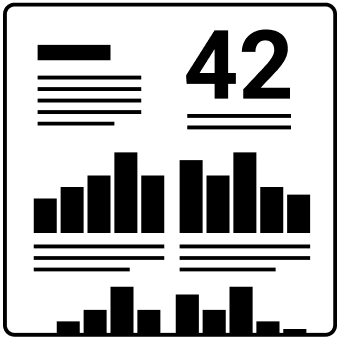}\textbf{Repository Dashboards (17\%)}---We found several examples that list a multitude of charts across multiple pages\linefig{pagination-multiple}, with \textit{overflow}\linefig{pagination-scroll} structure that impedes proper analytics, i.e., comparing views. 
The charts can be very detailed\linefig{data-complex}\linefig{data-visualization} and often lack any textual explanations, except for meta data information (which is often extensive). Charts may provide some interaction and usually provide links to explore\linefig{int-exploration}, drilldown\linefig{int-filter_focus}, and eventually download open data. Data and visualizations are updated regularly, while choosing very common \textit{visualizations} and \textit{numbers}, and parameterization\linefig{pagination-parameterization} to select data sets. Extensive \textit{meta information} is often provided for transparency. Examples include repositories from governmental and academic institutions, like Our-world-in data~\cite{roser2020coronavirus} or  Public Health Scotland (\#109). 

\parfig{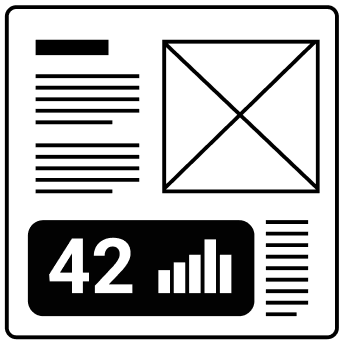}\textbf{Embedded Mini Dashboards (4\%)}---Some dashboards were found to be embedded into other applications such as news websites. These concise dashboards only occupied a small area on screen and usually come with concise visualizations (\linefig{data-signature}\linefig{data-number} \linefig{data-arrows}). Mini dashboards requires interactive features for navigation\linefig{int-navigation} and to parameterize the content quickly\linefig{pagination-parameterization}. \autoref{fig:pagination-tabs} shows an example of a mini dashboard embedded into a news page (\#83); like similar mini dashboards, it is \textit{linked} to a more in-depth \new{magazine} or \new{infographic} dashboard that invites further exploration.

\section{Discussing Design Tradeoffs}
\label{sec:tradeoffs}

There are many decisions to be considered in a dashboard design process, from high-level (e.g., selecting data and display devices) to more low-level decisions (e.g., color palettes, plot size, pagination structure). High-level decisions will almost always rely on sources over which the designer has little to no agency: the intended audience, the devices, and scenario in which a dashboard is used, or the wider team of data analysts and developers supporting the dashboard's creation. By providing specific solutions that seemed to have worked well in the past, our design patterns and dashboard genres can support lower-level design decisions that the dashboard designer actually has agency over and which they must solve to satisfy their requirements. These decisions include, e.g., the use of \textit{screenspace}, dashboard \textit{structure}, \textit{page layout}, \textit{color schemes}, \textit{visual representations}, etc. In this section, we reflect on our own design process in creating over 7000 dashboards for \covid-related data in the UK~\cite{khan2021propagating} as well as our discussions from both preparing and running a dashboard design workshop (\autoref{sec:workshop}). 

\subsection{Patterns to Balance Design Parameters}

From an information-theoretic perspective, a dashboard encodes a data space that is smaller than the data space of the data to be displayed. This requires a designer to decide which information to \textit{not show on screen} and how a user can then obtain the remaining information---if at all. One solution is to reduce information by \textit{abstracting} data and its visual encodings. Consider a time series representing the daily number of positive test cases for \covid during a period of 500 days. \autoref{fig:pagination-tabs} shows a design with four visual encodings (\textit{number}\linefig{data-number}, \textit{trend-arrow}\linefig{data-arrows}, \textit{miniature chart}\linefig{data-signature} and \textit{detailed visualization}\linefig{data-visualization}) for the \textit{same} time series data. They all lose information, but in different ways: the line chart visualization may lose information through its limited height, and the vertical pixel-resolution limits the range of data visible without scrolling. The large number shows the latest figure (a \textit{single-value}\linefig{data-simple}), while omitting all other data points in the time series dataset\linefig{data-complex}. The trend arrow (a \textit{derived value}\linefig{data-derived}) and \textit{miniature charts}\linefig{data-signature} are different levels of abstraction between the number and the detailed visualization, representing the full range of abstraction in one dashboard view.

However, information loss may also negatively cause confusion, misinterpretation, or erroneous judgment. At this moment, the dashboard designer needs to counterbalance this information loss through other means such as, e.g., adding interaction such as tooltips\linefig{pagination-tooltip}, scrolling\linefig{pagination-scroll} or spread content across multiple pages\linefig{pagination-multiple}. It is clear from this example that there are costs and benefits associated with different levels of abstraction and their visual encodings. However, the tradeoff here is the excessive cost of screen space when displaying several redundant visual encodings, and the increased cognitive cost of interpreting different levels of abstraction over the same data. It is necessary to display fewer numbers and visual information, to reduce the visual complexity of the dashboard. Designers thus need to find an optimal balance between abstraction and cost, e.g., to encode the latest data value(s) as numbers, and rely on perceptually less-precise plots to present overviews while reminding users of their past observations. With careful reasoning, exploring our dashboard design space does not necessarily involve exclusively considering many design options in a combinatory manner; rather to guide dashboard design in a considered combination of patterns.

\begin{figure}[h]
    \centering
    \includegraphics[width=1\columnwidth]{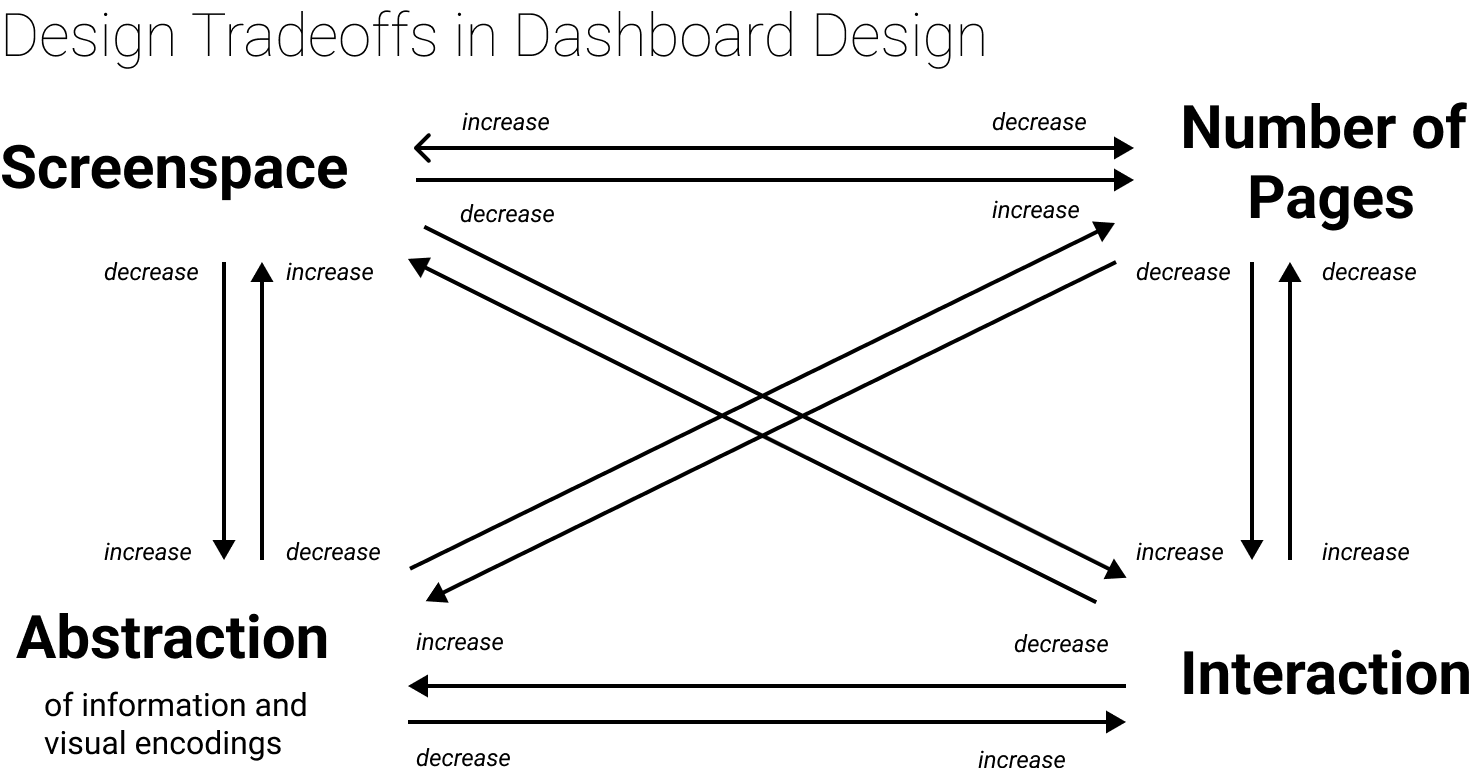}
    \caption{A simplified model for design tradeoffs in dashboard design. Reducing one of the parameters (screenspace, abstraction, number of pages, interaction) requires an increase in one of multiple of the other parameters.}
    \label{fig:tradeoffs}
\end{figure}

\autoref{fig:tradeoffs} illustrates a model for balancing possible \textit{parameters} in a dashboard designprocess. Such parameters can be, e.g., \textit{abstraction}, \textit{screenspace}, \textit{number of pages}, and \textit{interactivity}. In a design process, the goal is to \textit{minimize} each of these parameters. For example, to fit as much \textbf{information} as possible (low abstraction) into the least amount of \textbf{screenspace}, with no \textbf{interaction}, and to only show a \textbf{single page}. Such a solution would possibly be the gold standard in dashboard design: showing all important information at a glance, without the need for costly interaction. The model explains that negotiating tradeoffs is best represented through a \textit{stress} function between the parameters at hand. In the rest of this section, we discuss tradeoffs that minimize this stress with the help of design patterns.

Increasing the screen size is an obvious way to deal with potential information loss, as there is less need to abstract and reduce the amount of visible content. Where large screens are available, information loss can be minimized: e.g., dashboard \#42 spans the width of a room using multiple screens. However, large screens are not always available or practical, especially as dashboards find a more general audience who browse via personal and mobile devices.

Displaying a lot of information on a `typical' screen size requires careful \textit{page layout} and \textit{structure} of components and information; i.e., deciding how many pages\,\linefig{pagination-multiple} are necessary, what to show on each page, how to lay out those components. A good layout and structure needs to consider possible facets in data (e.g., vaccinations, cases, hospitalizations in the case of a \covid dashboard), as well as the tasks that require information \textit{across} these facets (e.g., comparing cases \textit{and} vaccinations). At the same time, a layout needs to prioritize information, e.g., showing information in different sizes or places on each page, perhaps using stratification\,\linefig{layout-strata} to put the most important information at the beginning.

When multiple facets and similar/repeated information need to be shown, \textit{table}\,\linefig{layout-grid} layouts may be ideal. Likewise, a layout can use repetition within each component, like that shown in \autoref{fig:layouts}--right, which uses a number\,\linefig{data-number}, trend-arrow\,\linefig{data-arrows}, and signature chart\,\linefig{data-signature} to convey data. Repetition like this guides a viewers' eye and helps interpretation and retrieval of information.
In contrast to structuring information across pages, dashboard designers can opt for simpler \textit{static dashboards}, which show data concisely on a \textit{screenfit}\,\linefig{pagination-screenfit} page without requiring user interaction. Static dashboards are ideal when interaction is not possible, desired, or necessary, and are also suited to print media. However, the trade-off associated with screen size is that static dashboards require carefully-chosen abstraction, to make the best use of available space (as discussed in the previous section).

Interactive \textit{parameterization}\,\linefig{pagination-parameterization} can provide balance between static and paginated dashboards: reducing the amount of information shown so that it fits on screen, but requiring the user to indicate what information is most relevant to their needs/tasks. Our design space encapsulates many interactive components that support \textit{exploration} \linefig{int-exploration}, (e.g., tooltips, filtering, search) and \textit{navigation} \linefig{int-navigation}, through multiple pages of information (e.g., search, tabs, links). Interaction gives viewers the ability to \textit{personalize} \linefig{int-personalization} and use the dashboards in a way that suits their needs. Contrast this with static dashboards that are framed by designers, revealing strong parallels to author vs. reader-driven storytelling~\cite{segel2010narrative}.

Especially for narrative examples of dashboards (magazine dashboards\,\linefig{../type-magazine}, inforgraphic dashboards\,\linefig{../type-infographic}) interaction can help set the narrative structure and pace. Interaction can be utilized to streamline the volume/velocity of information communicated, i.e., slowly revealing parts of the information/data based on interaction. The most simple interactions to deal with more information than can fit on screen are scrolling (in an \textit{overflow} layout\,\linefig{pagination-scroll}) or navigation buttons\,\linefig{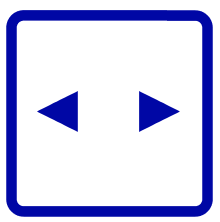}. Those interactions do not interfere with the individual visual encodings (i.e., visual encodings can be static images) and they allow designers to create dashboards that are easily responsive across different screensizes. Other simple options that do not interfere with the implementation of visual encodings include use of \textit{tabs}\,\linefig{pagination-tabs} and \textit{links}\,\linefig{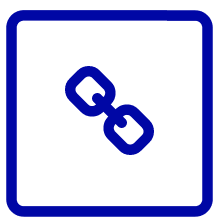}. Detail-on-demand \linefig{pagination-tooltip} and parameterization \linefig{pagination-parameterization} may require a more specific implementation, but can be effective, as discussed earlier.

\subsection{Curated Dashboards vs. Data Collections}
\label{sec:tradeoffs-templates}

The \numtemplatestext dashboard genres presented in \autoref{sec:types} represent different ways that dashboard design decisions and patterns can be combined to create usable dashboards, typically oriented towards specific contexts, tasks, and users. In this section, we discuss differences and tradeoffs between these, to inform their choice in future dashboard design.

We see a distinction between \textbf{curated dashboards} (static dashboard \linefig{../type-classic}, magazine dashboard \linefig{../type-magazine}, infographic dashboard \linefig{../type-infographic}) that are highly selective of data and visual representation towards a specific goal, and \textbf{data collections} (analytic dashboard \linefig{../type-analytic}, repository dashboard \linefig{../type-repository}) that aim to transmit large volumes of information so that viewers can seek the information most relevant to them. Curated dashboards can be described as author-driven storytelling, while collection dashboards can be described as reader-driven storytelling~\cite{segel2010narrative}.

\textbf{Curated dashboards} will typically fit on a single screen with a \textit{screenfit}\,\linefig{pagination-screenfit} or \textit{overflow}\,\linefig{pagination-scroll} layout, and offer limited ability to change the data or its visual encodings (i.e., limited \textit{parameterization}\,\linefig{pagination-parameterization}). Creating such dashboards requires a greater extent of curation, design, and selection, i.e., representing \textit{the right information} using \textit{the right visual encodings}. These dashboards will have well-defined use cases, and the designer's role is to `translate' the data for that purpose~\cite{Kitchin2015urbandashboards}.

\textbf{Data collections} are dashboards that provide access to lots of data, represented in different ways for different tasks (e.g., for analytical use or for sharing open data). Although a degree of curation is required (i.e., the designer has made deliberate decisions about \textit{how} to show the data), there is less need to reduce the amount of visible content. These dashboards typically use \textit{pagination}\,\linefig{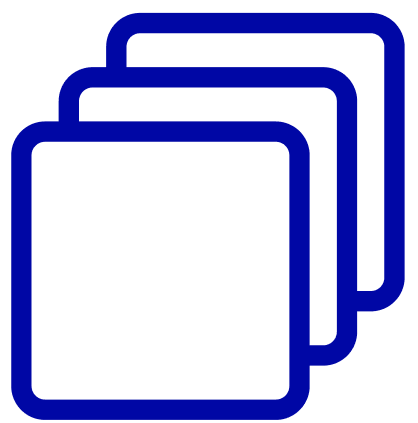} and/or \textit{overflow}\,\linefig{pagination-scroll}, as the goal is not to make the most economic use of screen space but to maximize the amount of information available to users, so that a viewer can quickly find what they are most interested in. The use cases for data collection dashboards are more open-ended and they are often aimed at a more broad audience. 
\section{Dashboard Design Workshop}
\label{sec:workshop}

We ran an online dashboard design workshop to help us scrutinize and refine the design pattern collection as well as our discussion on design tradeoffs. \new{In the workshop, participants applied the patterns to their own dashboard projects.} The workshop ran for 2 weeks and was open to everyone with a dashboard design task at hand. It aimed to guide people through the main stages and decisions of a design process while using the design patterns for ideation and discussion. Rather than implementing a dashboard in a tool such as Tableau or Power BI, the course intended to finish with interactive mockups in Figma. 
Detailed information as well as all material from the workshop can be found online: \url{https://dashboarddesignpatterns.github.io}.

\subsection{Setup and Participants}
\label{sec:setup}

\textbf{Outline:} A (1) \textbf{3h kickoff session} gave a fast-forward through the design process and introduced the design patterns. The session started with an 
(1.1) overview of example dashboards from our collection, asking participants to describe what they see and how these dashboards appear to them upon first sight. Then, we
(1.2) introduced dashboard design challenges and discussed high-level guidelines such as mentioned in Section 2. Then, 
(1.3) the workshop introduced the framework in Figure \ref{fig:tradeoffs}. Eventually, 
(1.4) the session introduced a possible design process, one stage for each type of design patterns:
(i) \textit{Data Information} (what information about data do I need to show)
(ii) \textit{Dashboard structure} (which pages to I need, how are they related?),
(iii) page design (\textit{page fit} and \textit{layout}, \textit{visual representation}), and eventually 
(iv) \textit{interactivity} and (v) \textit{color}. The full process is outlined on our website.\footnote{\url{https://dashboarddesignpatterns.github.io/processguidelines.html}}
Following the kickoff session, we had \textbf{(2) six asynchronous check-in sessions} over two weeks. Each session lasted 1h and was facilitated by one to two of the co-authors. During these sessions, we provided individual help to participants, discussing their dashboard examples, giving advice about pattern usage and discussing related data visualization questions. Each session was attended by six participants on average, discussing 2-4 dashboards during that hour. Eventually, we had a \textbf{(3) debriefing meeting} in which we asked participants to show their mockups and reflect on their design process, choices, and challenges. A post-hoc questionnaire collected feedback about the workshop outline and design patterns. The entire workshop was held on Zoom, encouraging participants to work on their design mockups asynchronously and share their designs through Figma.

\textbf{Participants:} We had 23 participants in total from diverse backgrounds, including data science, medicine, psychology, economics, bioinformatics, design, etc., including a mix of industry and academic partners. Each participant worked on their own dashboard within a context they defined themselves and which was relevant to their work. Example dashboards created during the course included an analytic dashboard allowing to browse learning analytics for hundreds of university courses; an infographic dashboard to inform midwives about the risks of \covid{} and vaccination during pregnancy; an analytic dashboard to track dynamic user interaction log data; a repository dashboard tracking live data about conflicts across the world; an analytical dashboard to help modelers assess and track the performance of their models over various sessions of training and parameter adjustments; a repository dashboard about open government data to allow journalists and politicians retrieve relevant data; and eventually a personal analytical dashboard to monitor energy consumption at homes. About half of the participants had prior experience of one to several years in dashboard design with Power BI and Tableau. For almost all participants, this workshops was the first time they deliberately engaged in dashboard design in a structured and guided form.

\newcommand{\says}[1]{\textit{``#1''}}

\subsection{Findings}
\subsubsection{Workshop Discussions}

In the drop-in sessions, discussions were sparked by participants sharing their designs. Participants also shared questions, possible answers, and reflections, voiced in a participatory manner. In the following, we summarize key \new{discussion points and the role of design tradeoffs and design patterns in this discussion.} 

\textbf{Information overload} was a frequent topic related to \textit{what} information to include in a dashboard and \textit{how much} information to include. \new{When discussing dashboard examples at the beginning of the course, many} participants complained about \says{too much info}, data, and information, and general confusion about where to look at. Some commented on small \new{sizes of} charts and maps, and unclear relations between components: \says{not sure how the linechart relates to the number}. Color was sometimes considered overused and its particular function was not always clear. \new{Our color patterns helped sensitizing participants here and discuss reasons for using color. } 

\textbf{Optimizing screenspace and reducing the number of pages} was another common topic. While discussing techniques to make visualizations simple, concise, and make best use of the screen space, one participant asked \says{can a dashboard ever contain too much information?} This comment sparked a long discussion around how to best \new{apply and combine} \textit{screenspace} patterns \new{(screenfit, overflow, detail-on-demand, parameterization, multiple pages)}, layouts \new{(open, table, stratified, grouped, schematic)}, and compressed visual encodings \new{(namely miniature charts and single values). The discussion} resulted in some agreement that there are lots of techniques to visually compress dashboards \new{and that, again, the design patterns provided an excellent repertoire of potential solutions (also see 6.2.3)}. One participant raised the question how to design for mobile and desktop screens and how many versions of a single dashboard are required. \new{We discussed how our patterns can help reducing a large onscreen dashboard into a mobile one, by applying a single pattern (e.g., multiple pages) or gradually applying several patterns to the large onscreen dashboard.} 

These issues eventually led to discussion about \textbf{novice vs.\,expert audiences} as well as \textbf{casual vs.\,frequent users}. Novice or casual users will likely need more guidance in the form of clear layout \new{(e.g, stratified)} and titles, little to no interaction, and generally less information \new{(e.g., single value, aggregated data)}. Expert or frequent users will be more data literate and require more data for their decisions. At the same time, dashboards for experts and frequent users will also likely include more bespoke features, compared to dashboards for novices and casual users, who might be driven by general concerns and more general tasks. Novice and casual users will likely appreciate more message-heavy dashboards (e.g., infographic dashboards\,\linefig{../type-infographic} or magazine dashboards \linefig{../type-magazine}), while experts might appreciate more \says{self-serving} dashboards, e.g., analytical\,\linefig{../type-analytic} or repositories\,\linefig{../type-repository}.

Many dashboard examples we discussed were found to suffer from a \textbf{lack of context}. For example, individual values \linefig{data-simple} and numbers \linefig{data-number} require more data to be compared to. Such data can be provided through a \textit{temporal context}, e.g., data from the past, or other \textit{comparisons} (e.g., miniature charts, detailed data, trend arrows, indicators). Experts and frequent users might have an implicit understanding of this context, but novices and casual users might not. 

\textbf{Color and accessibility} was a frequent topic in our discussions. Color was heavily used in the dashboard designs, either originating from the participants' personal decision or provided by the dashboard tool without consultation by the designer. Color consistency was a big issue in most designs and most advice hinted towards removing color where not necessary, e.g., on most \new{screenfit} dashboards. Accessibility was raised as an often neglected topic with implications for titles, colors, and chart sizes. While many guidelines exist for accessibility in visualization, we are not aware of any accessibility guidelines that take into account the density, tasks, and scenarios of dashboards.

Those participants who had some \textbf{experience with dashboard design tools} commented that design iterations are very easy and building dashboard prototypes was easy. On the downside, these tools offered complementary support for dashboard creation with no tool clearly outpacing the other tools or supporting all design demands \new{and design patterns}. Many tools lacked support for specific design decisions and \new{design patterns (e.g., optimal use of screen space, layout, interactions, shared color scheme\linefig{color-distinct}}).

\newcommand{\saidby}[2]{\says{#1}{\small[#2]}}
\newcommand{\pbobby}{P1}
\newcommand{\pmed}{P2}
\newcommand{\ptwo}{P3}
\newcommand{\pgtwo}{P4}
\newcommand{\pgthree}{P5}

\subsubsection{Reflections on the Design Process}

We obtained many quotes from our questionnaire which nicely capture participants' reflections and which do not need much commenting:

\begin{itemize}[noitemsep,leftmargin=*]
    \item \saidby{Start with simple designs that spotlight just the most important data, then add to this as you go; this will help avoid just putting everything you can on the dashboard, keeping it focused. (This might not apply to repository style dashboards).}{\pgtwo}
    \item \saidby{I learned not to try to ask too many questions for one graph. One graph can give several different answers, but the question should just be one question.}{\ptwo}
    \item \saidby{Simpler is often better when it comes to charts: bar charts, line charts, and tables are often clearer and easier to read than their more fancy looking counterparts.}{\pgtwo}
    \item \saidby{If you try to simplify [your design] too much, you risk imposing your own story.}{\ptwo}
    \item \saidby{Use filters, menus, buttons, and parameters to reduce the number of static visuals shown at any one time [to] keep things readable.}{\pgtwo}
    \item \saidby{In medicine we often hand out 15 leaflets and tell patients [...] to come back with a decision [...] but I have not given them information about how to \emph{use} [the information] and how to \emph{decide}.}{\pmed}
\end{itemize}

\subsubsection{Pattern and Workshop Utility}

Participants valued the individual feedback onto their designs as well as discussing other participants' design. They also valued the framework, guidance, and community the workshop provided. 

\begin{itemize}[noitemsep,leftmargin=*]
    \item \saidby{[Design patterns and their terminology] limit down [the design complexity] from endless design options to 2-3 good candidates [...] I think it will be really useful for co-creation or kick-off meetings [...] so everyone is very clear on what the key elements are and what we're trying to achieve right from the start etc}{\pbobby}
    \item \saidby{[I used the patterns] first as guidance for designing [then] I did any changes with the design principles in mind. In the end, I used [the patterns] as a `check-list' to review my design decisions.}{\pgthree}
    \item \saidby{If you walk through each of those patterns/categories, you basically end up with a written plan for your dashboard which is clear to everyone e.g. `I'm planning to build an analytic dashboard. Because of XYZ factors, we're going to use a paginated design and allow user interaction etc.'}{\pbobby}
    \item \saidby{I think it's very useful to now be able to ask things like `is this a static dashboard? is this an analytic dashboard?'. Or on structure, things like 'is a hierarchical structure the best? [...] it makes it a bit easier to envision and make choices.}{\pbobby}
    \item \saidby{in my experience as a data analyst, building dashboards is something I've had to pick up and learn myself with not much outside training or workshops [...], that's a somewhat common experience for analysts.}{\pbobby}
 \end{itemize}
\section{Towards a Design Framework for Dashboards}
\label{sec:discussion}

At this point, we want to reflect on how the insights from this research can be put into actionable dashboard design practice. To that end, we consider a design framework consisting of four parts: (1)\,\textit{knowledge} in the form of guidelines, annotated examples, and design patterns; (2)\,\textit{tradeoffs} that require discussion and decision making; (3)\,\textit{design processes}; and (4)\,\textit{tool support}. Some of these aspects can be drawn from the literature (e.g., knowledge as discussed in \autoref{sec:relwork}); some are presented in this paper (e.g., design patterns, annotated exaples, tradeoffs, design processes); others are compelling topics for future work and can be informed by our findings here (e.g., tool support for effective dashboard creation).

\subsection{Design Knowledge}
\label{sec:discussion-knowledge}

Design knowledge includes anything that can be learned and read from the literature, or taught in classes in the form of \textit{theory}. We can think of this as the `raw material' in any design process: the ideas, theories, models and examples that compose the syntactic and structural rules and components of any design. Design knowledge can come from empirical observation from studies, and is generally highly formalized and includes guidelines, examples, and design patterns.

Design patterns for dashboards are, for the first time, described in this paper. We also provide a set of dashboard \types that illustrate common combinations of design patterns and abstract many of the real-world examples examined in our work. These \types can be put into action as part of the dashboard design process. However, we believe more design patterns could be found and our collection paves the way to a more comprehensive collection.
These patterns help establish a terminology for design knowledge about dashboards. They might even point to specific characteristics of dashboards, helping to see dashboards as only an \textit{instance} within the forms and genres in the wider visualization landscape, such as multiple coordinated views or infographics. At the same time, these patterns help relate dashboards to these other forms and genres, while keeping the delineation of dashboards deliberately blurred to allow for transfer and cross-fertilization of design ideas.

An additional contribution is our dashboard corpus. We extended the dashboard collection started by Sarikaya et al.~\cite{Sarikaya2019Dashboards} to a set of \numdashboards examples. Due to the sheer number of dashboards online, any systematic survey is impossible, but a common corpus can inform  
future research, design, and discussion, similar to examples for physical data visualization~\cite{physlist}, storytelling~\cite{bach2018narrative}, or data comics~\cite{bach2018design} that involve plenty of examples from outside academia.
Dashboard design guidelines were discussed in \autoref{sec:relwork-guidelines} and applied and scrutinized through our workshop. Guidelines can often be generic and widely applicable, but most are limited to specific use cases and some guidelines may contradict each other. Based on our design patterns and discussion of tradeoffs, we are able to formulate additional guidelines, listed on our website.

\subsection{Design Tradeoffs}
\label{sec:discussion-tradeoffs}

Every design process and design problem is unique in that several parameters must be considered: users, tasks, contexts, devices, etc. Design tradeoffs are inevitable when no solution is optimal, i.e., when the specific parameters of a design problem have contradictory \textit{knowledge} (e.g., guidelines, heuristics, solutions). Dashboard designers can use this knowledge to inform their approach, but other activities (deliberate or otherwise) will be necessary: e.g., reasoning and logic, experimentation and prototyping, user-centered design and evaluation. Decisions may likely influence or conflict with other decisions, causing further design tradeoffs to be necessary, requiring constant iteration towards an effective and usable dashboard design.

We discussed a framework for design tradeoffs (\autoref{sec:tradeoffs}), strongly informed by our design patterns and by our experience in designing dashboards~\cite{khan2021propagating, Khan2022TSC}. We think of this discussion as a first formal discussion---partially based on information theory---of design decisions for dashboards. This discourse will evolve with the set of design patterns and future empirical studies. Future empirical studies should challenge our reflections and address the specific design tensions described in this paper. There are also many questions left open regarding the costs and benefits of specific solutions and decisions.
For example,
\textit{To what extent do users engage with interactive dashboards?} (some similar work exists in the context of interactivity and storytelling on the web~\cite{boy2015storytelling}); 
\textit{Does interaction help users solve tasks effectively and efficiently, or does it just add complexity?};
\textit{How much data information is too much?} \textit{How to support specific tasks and audiences?}

\subsection{Dashboard Design Processes}
\label{sec:discussion-process}

There is a key need for dashboard design processes that structure both knowledge and decisions in tradeoffs, to guide designers towards effective dashboard designs. As part of our workshop, we describe one possible design process (see website) that aims to describe mid-level decisions and complement the discussion in \autoref{sec:tradeoffs}. The process assumes requirement analysis has defined \textit{users}, \textit{tasks}, and \textit{datasets}. This process is intended to kickstart the dashboard design process and introduce our design patterns one group at a time. The process implies many iterations to negotiate design tradeoffs.

\subsection{Tool and Technical Support}
\label{sec:discussion-tools}

While there are existing tools for dashboard creation (e.g., Tableau, Power BI, Exploration Views~\cite{elias2011exploration}, QualDash~\cite{elshehaly2020qualdash}), there is a need for greater support to guide design choices. For example, authors of dashboards and dashboard-authoring tool designers could offer support for a range of design patterns and genre templates. This could streamline dashboard design and allow designers to create different dashboard versions based on the same widgets. There might be further potential for automating dashboard creation~\cite{key2012vizdeck} through recommender systems, e.g., as used with visualizations~\cite{Law2020_insights} and infographics~\cite{wang2019datashot}. Recommender systems should take into account levels of abstraction and composition. However, these are non-trivial challenges that require formal design rules, which in turn require further study.

Dashboard users could be provided with options for personalization, sharing, bookmarking, or annotation. Personalization could allow a user to specify data they consider important and adapt the layout and size of components accordingly. Only a small number of dashboards support personalization through adding or moving/resizing widgets. A related problem is a lack of responsiveness to different screen sizes. There has been only a handful of studies and suggestions how to make visualizations responsive~\cite{hoffswell2020techniques} but responsiveness in dashboards should include the levels of abstraction for data and visual encodings, eventually turning a \textit{static dashboard} \linefig{../type-classic} into an \textit{analytic dashboard} \linefig{../type-analytic}.

\subsection{Limitations}

Our research is limited by the set of dashboards we have curated. To ensure consistency with existing research, we started our collection from existing dashboard used by Sarikaya et al.~\cite{Sarikaya2019Dashboards} and added contemporary dashboards, mostly around Covid. We open our dashboard collection and coding scheme for future research on our website. Our design patterns reflect close agreement among six coders and have been scrutinized through application in the workshop. We highlight that any design pattern collection (\autoref{sec:relwork-patterns}) is highly \textit{qualitative} in that it aims \textit{ideation} and describing applicable solutions for reuse, inspiration, and analysis. This implies patterns to bear meaning and capture ideas that can be applied. We see our patterns allowing hybrids, e.g., parallel structures in hierarchical structures, or hybrids of grouped and stratified and open layouts. Our patterns and their frequencies (reported as \% in \autoref{sec:patterns}) are representative for the dashboards we analyzed. Future work can complement our collection. Eventually, the scope of this paper is not to suggest `good designs’, but to inform future studies.
\section{Conclusion}

\new{Dashboards have their distinct place in the visualization landscape, yet are still an overlooked subject in the VIS community. They share many characteristics with, e.g., multiple coordinated views, small multiples, infographics, and potentially data comics but lack generic and practical design guidance beyond high-level guidelines. Our dashboard patterns are the first to map common solutions in dashboard design in an applicable way. They help describing why dashboards are special: abstraction of data, organizing of a screenspace, grouping of elements, showing relations (hierarchy, grouping, color), and the use of interaction for exploration, drilldown, navigation, or personalization. However, we argue against a strict definition of the term `dashboard', and rather see it as way to communicate specific affordances onto a visualization user interface in a specific context: the need for overview, control, and conciseness for decision making.}

\new{In proposing a simple model for discussing design tradeoffs, we show how the design patterns can be used in a deliberate and informed design practice. Our dashboard genres describe higher level patterns and solutions to design tradeoffs and target solutions for higher level questions such as audience, extent of data, usage scenarios, and medium (e.g., static vs. interactive, mobile phone vs. large wall). Patterns, tradeoffs, and genres are meant as a first stepping stone that provides structured guidance (terminology, examples, ideas) to both novice and expert users access to dashboard design. We are aware of the limitations of our classification and tradeoff framework and the necessary unfinished nature of any such current or future attempt. It was beyond the scope of this work to determine what makes for a `good' dashboard, and this will depend on many subjective and domain-specific factors. The diversity in dashboard design found here suggests more research into strengths and weaknesses of different dashboard genres, echoing Sarikaya et al.'s call-to-action for more dashboard visualization research~\cite{Sarikaya2019Dashboards}.}

\acknowledgments{\textit{This work was supported by EPSRC (EP/V054236/1). We would like to thank
all volunteers from the SCRC and all VIS volunteers~\cite{khan2021propagating} and the dashboard workshop. In particular, we would like to thank Dr. P. H. Nguyen (Red
Sift Ltd.) and Dr. H. C. Purchase (U. Glasgow) for their contribution to the RAMPVIS work on dashboards, and SCRC colleagues Profs. R. Reeve (U. Glasgow) and L. Matthews (U. Glasgow) for their advice on pandemic responses.}}

\bibliographystyle{abbrv-doi}
\bibliography{main}

\end{document}